\documentclass[pre,twocolumn,showpacs]{revtex4}

\usepackage{graphicx}
\usepackage{bm}

\newcommand{\avec}[1]{{\bm{#1}}}
\newcommand{\avecu}[1]{{\hat{\bm{#1}}}}
\newcommand{\avect}[1]{{\tilde{\bm{#1}}}}
\newcommand{\amat}[1]{\bm{#1}}
\newcommand{\tr}[1]{#1^{\scriptscriptstyle T}}
\newcommand{\spa}[1]{{\rm #1}}

\begin{document}

\title{Molecular dynamics simulation of polymer helix formation using rigid-link
methods}

\author{D. C. Rapaport}
\email{rapaport@mail.biu.ac.il}
\affiliation{Physics Department, Bar-Ilan University, Ramat-Gan 52900, Israel}

\date{February 14, 2002}

\begin{abstract}

Molecular dynamics simulations are used to study structure formation in simple
model polymer chains that are subject to excluded volume and torsional
interactions. The changing conformations exhibited by chains of different lengths
under gradual cooling are followed until each reaches a state from which no further
change is possible. The interactions are chosen so that the true ground state is a
helix, and a high proportion of simulation runs succeed in reaching this state; the
fraction that manage to form defect-free helices is a function of both chain length
and cooling rate. In order to demonstrate behavior analogous to the formation of
protein tertiary structure, additional attractive interactions are introduced into
the model, leading to the appearance of aligned, antiparallel helix pairs. The
simulations employ a computational approach that deals directly with the internal
coordinates in a recursive manner; this representation is able to maintain constant
bond lengths and angles without the necessity of treating them as an algebraic
constraint problem supplementary to the equations of motion.

\end{abstract}


\maketitle

\section{Introduction}

Polymers, because of their importance and complexity, have provided a longstanding
challenge for computer simulation. Over the years, the field has become fragmented,
both in terms of the problems addressed and the methodology employed. Broadly
speaking, the kinds of system studied can be classified into distinct groups; there
are biological heteropolymers, a category dominated by the proteins; homopolymers
and block copolymers that include a great variety of molecular types, from alkanes
to plastics; and idealized polymer models used for elucidating general principles
such as the theta point, reptation, and multiphase behavior. The computational
techniques span an equally broad range; they include molecular dynamics (MD)
simulation employing models that represent the molecules at various levels of
detail, ranging from fully atomic to highly reduced descriptions; Monte Carlo
sampling of both continuum- and lattice-based systems, again with different levels
of representation; and exact enumeration of small systems aimed at eliminating the
sampling errors inherent in the other methods. While all three kinds of methodology
provide important information about equilibrium behavior and, in a sense, amount to
doing statistical mechanics numerically, it is only the MD approach that provides
access to the dynamical and nonequilibrium aspects of the behavior; while it might
be argued that Monte Carlo shares some of this capability, the associated dynamics
is artificial and entirely a consequence of the chosen stochastic sampling
algorithm, and so bears little relationship to Newtonian dynamics. Lattice-based
approaches, though offering a vastly reduced configuration space, have the
additional problem of the discreteness of the lattice on which the polymer is
embedded, and the consequent absence of gradual transitions between different
configurations.

The inherent difficulty in polymer simulation is that the problem naturally
embraces a broad range of timescales, ranging from very fast processes associated
with bond vibration, followed by the somewhat slower, highly localized
conformational changes such as crankshaft motions, then the even slower aspects of
reorganization such as the still relatively localized process of helix formation,
and, finally, the typically extremely slow changes that lead to the emergence of
tertiary structure characteristic of protein folding and to polymer diffusion in a
concentrated solution. The timescales associated with this hierarchy of processes
span a range considerably in excess of ten orders of magnitude, and so such systems
are clearly not generally amenable to direct modeling, unless subjected to major
simplification. Considerable effort has been invested in the design of models and
simulation methods with the aim of alleviating this problem to at least some
degree.

One especially important application of polymer simulation is in the field of
protein folding, e.g., \cite{bro88,dil95,sha97,bro98,pan00}; achieving an
understanding of the mechanisms underlying this important process presents a major
challenge to computational biochemistry. Protein modeling runs the gamut from, at
one extreme, highly detailed molecular representations involving potentials derived
from a mixture of theory and experiment, together with a solvent of individual
water molecules, all solved by MD and an enormous amount of computational effort 
\cite{dua98,ber98}, through highly simplified models also solved by MD
\cite{lev75}, to yet even simpler models embedded in lattices with only a limited
number of degrees of freedom (DOFs) studied using a suitable Monte Carlo procedure
and a greatly reduced investment in computing \cite{dil95}; even complete
enumeration of all conformations is sometimes feasible \cite{cri98}. While the
manner in which the amino acid sequence of any given protein is able to determine
its presumably unique spatial structure continues to be the subject of intense
study, of no less importance is the question of the folding pathway -- the
preferred route (or routes) through multidimensional conformation space eventually
terminating at the native state. While all the widely differing methodologies
enumerated above can be used for studying folded states, the collective dynamical
processes that underlie folding really demand an approach based on MD. But, after
reaching this conclusion, there is a practical question of whether, even after
substantial simplification, serious progress in understanding the mechanisms of
folding can be achieved by computer simulation, owing to the diversity of intrinsic
timescales; while substantial advances have been made, a great deal remains to be
done before this question is answered.

\begin{figure}
\includegraphics[scale=0.38]{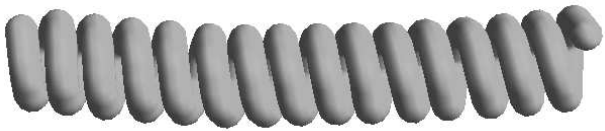}
\caption{\label{fig:lk3d_01}A well-formed helix in a chain of length 90; a goal of
the simulations is to observe chains spontaneously collapsing into this state (the
polymer is drawn as a tube whose radius is that of the monomers).}
\end{figure}

\begin{figure}
\includegraphics[scale=0.54]{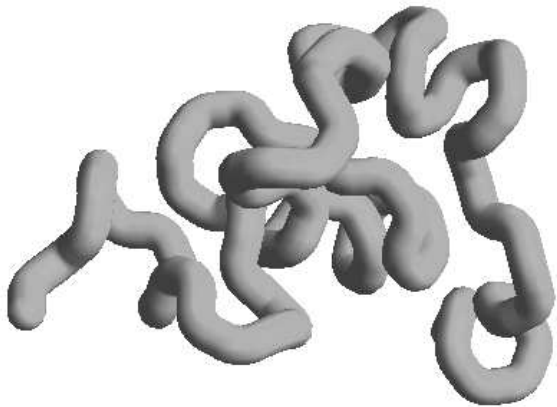}
\caption{\label{fig:lk3d_01a}A randomly coiled chain of length 90; this
configuration represents a typical state of the chain prior to the onset of
folding.}
\end{figure}

The goal of the present paper is twofold. The first goal is a demonstration of a
different perspective on the MD approach to studying protein folding. The most
ambitious level of modeling is based on carefully constructed potential functions,
often with a multitude of parameters; since the native conformation generally
corresponds to the state of minimum free energy, establishing the details of these
interatomic interactions, including solvent effects, provides the foundation for
such work. Determining whether the known native state of a given protein is the one
favored by energetic considerations is in itself a complex optimization task, but
following the full dynamics over a sufficiently long period of time for the major
structural changes that typify protein folding to occur verges on the impossible.
The approach adopted here is just the opposite, and the question posed is the
following: Given a known structural motif, such as the helix, and a simplified
model of a polymer chain with a readily determined, unique ground state
corresponding to this configuration, as in Fig.~\ref{fig:lk3d_01}, will the chain
collapse into this state within a reasonable amount of computation time when
allowed to move freely in space, as shown in Fig.~\ref{fig:lk3d_01a}, while
subjected to gradual cooling? 

The most elementary of these organized structures is the helix, which, while being
a prominent feature in many globular proteins, is only classified as a secondary
structural element (the primary structure being the amino acid sequence itself),
and because of its homogeneous nature (except for the ends) it might be argued that
being able to fold a helix is not really a significant step in learning how to fold
an entire protein. Therefore, another folding problem considered here is one with a
ground state formed from an antiparallel pair of helices. This, too, is a
recognizable element in some proteins, and is unquestionably classified as tertiary
structure.

The obvious extension of this approach, a subject for future exploration, is to
design simple models for other structural motifs, in the hope of learning more
about folding by examining the collapse pathways of these idealized models; some
structures might fold more readily than others, in which case the steric and
topological issues involved could be investigated; for some structures there might
be recognizable intermediate states along the folding trajectory; some cases might
reveal useful properties that, when regarded as conformational (or reaction)
``coordinates'', might serve in the design of other kinds of simplified models
\cite{cri98}; and finally, once the simple version has been found to have the
correct behavior, the models could be enhanced by gradually incorporating features
from more realistic representations, including specific interactions and structural
details. This represents the motivation for this kind of modeling approach.

The second goal is methodological. Even when considering the simplest of model
polymers, in which, typically, all the molecular detail is absorbed into effective
atoms located along the backbone chain (more so if this simplification is not made)
there is a need to specify the internal DOFs of the system. One possibility is to
assume that adjacent atoms are connected by stiff springs represented by a suitable
potential function; in this case each atom has its full complement of three
translational DOFs and, if these atoms are regarded as rigid particles rather than
point masses, three rotational DOFs as well. If the bond potentials are made
sufficiently stiff to correspond to a typical real system, the ensuing
high-frequency vibrations impose a very small integration timestep, which runs
contrary to the goal of efficiently simulating over long periods of time.

It is, however, possible to introduce geometrical restrictions, such as strictly
constant bond lengths, while retaining a soluble dynamical problem. This is done by
introducing holonomic constraints and Lagrange multipliers into the equations of
motion \cite{gol80}, and then solving a set of algebraic equations while
integrating the differential equations of motion. Two approaches have been
developed for doing this; one involves initially solving the unconstrained
equations of motion over a single timestep and then iteratively correcting the
relative coordinates \cite{ryc77,cic82}, and, optionally, also the relative
velocities \cite{and83}, using a relaxation procedure to ensure the constraints
remain satisfied; the other tackles the problem by constructing a matrix
representing the contributions of the constraints which, in effect, must be
inverted at each timestep \cite{edb86,rap95}, and which is subject to gradual drift
requiring regular correction. Similar geometric constraints can be introduced to
maintain constant bond angles as well, since it is often a reasonable approximation
to assume that the angles between consecutive bonds along the backbone (or
elsewhere) are unvarying. Such geometrical constraints have proved extremely
useful, given the nature of the excitations present in the system: fluctuations in
bond lengths, and sometimes also angles, tend to be of relatively small amplitude
and high frequency, so that freezing them out of the dynamics permits a substantial
increase in the allowed integration timestep. The amount of additional processing
required for the constraints depends on their number $n_c$; the dependence is
typically $O(n_c)$ for the iterative approach, but for the matrix approach it is
$O(n_c^3)$, making the latter unsuitable for large problems.

If bonds lengths and angles are fixed, the only remaining internal DOFs are the
dihedral angles, each defined in terms of a rotation about an axis lying along a
bond, and affecting the relative orientation of the pair of bonds on either side.
For reasons shrouded in history, dealing with this problem has been perceived as
difficult, as indeed it is, if the problem is not addressed in a suitable manner. A
significant advance in the methodology for dealing with dynamical problems
involving internal coordinates occurred some years ago in the robotics field
\cite{rod92,jai91}, but with only the occasional exception, e.g., \cite{bvc98}, it
appears to have gone unappreciated by the polymer simulation community at large.
Because of the importance of this technique, the goal of which is to deal directly
and economically with the internal DOFs, and since there is no reason why it should
not be capable of replacing the various constraint-based approaches for most
applications, a detailed treatment of the underlying theory is included in the
paper. 

This approach to the dynamics of linked bodies also requires solving the dynamics
of individual rigid bodies. An alternative, recently described means \cite{dul97}
of numerically dealing with the rigid-body equations of motion is discussed
briefly; the method is based on rotation matrices, rather than on quaternions (or
even Euler angles) that are generally used. The present formulation differs
slightly from the original in regard to the reference frame in which the
computations are carried out. The use of rotation matrices offers improved
numerical stability, and since the method belongs to the leapfrog family of
integrators, it means that simple leapfrog integration techniques can be used for
the entire set of dynamical equations appearing in the problem.

\section{Linked-body dynamics}

\subsection{Chain coordinates}

Consider a linear polymer chain whose monomers are joined by rigid bonds. In the
discussion that follows, the terms ``monomer'', ``atom'', ``site'' and ``joint''
will be used interchangeably, as appropriate to the context, likewise ``link'' and
``bond''. Bond lengths and angles are constant. If each torsional DOF is regarded
as a mechanical joint associated with the site at one end of the link, with just a
single rotational DOF, then the system is analogous to a basic problem in the field
of robotic manipulators \cite{rod92,jai91}.

The chain configuration is defined by the site positions $\{\avec{r}_k\}$, and if
the bond vectors between adjacent sites are $\{\avec{b}_k\}$ then $\avec{r}_{k+1} =
\avec{r}_k + \avec{b}_k$. The internal configuration of the chain can be specified
by a set of bond rotation matrices $\{\amat{R}_k\}$. The transformation between the
local coordinate frames attached to bonds $k-1$ and $k$ ($k \ge 1$) involves a
rotation through the bond angle $\alpha_k$ about the axis $\avecu{x}_{k-1}$, where
$\cos \alpha_k = \avecu{b}_{k-1} \cdot \avecu{b}_k$, followed by a rotation through
the dihedral angle $\theta_k$ about the joint axis $\avecu{z}_{k-1}$. The matrix
(actually its transpose) corresponding to this rotation is 
\begin{equation}
\amat{R}^T_{k-1,k} = 
\left(\begin{array}{ccc}
\cos \theta_k & - \sin \theta_k \cos \alpha_k &
\hphantom{-} \sin \theta_k \sin \alpha_k \\
\sin \theta_k & \hphantom{-} \cos \theta_k \cos \alpha_k &
- \cos \theta_k \sin \alpha_k \\
0 &\hphantom{-} \sin \alpha_k & \hphantom{-} \cos \alpha_k
\end{array}\right) , \label{eq:rotmat}
\end{equation}
so that
\begin{equation}
\tr{\amat{R}}_k = \tr{\amat{R}}_0 \tr{\amat{R}}_{0,1} \cdots \tr{\amat{R}}_{k-1,k} ,
\end{equation}
where $\tr{\amat{R}}_0$ represents the orientation of the initial site and bond,
and
\begin{equation}
\avec{r}_{k+1} = \avec{r}_k + |\avec{b}_k| \tr{\amat{R}}_k \avecu{z} .
\label{eq:sitecoord}
\end{equation}

In the present case, $\{|\avec{b}_k|\}$ and $\{\alpha_k\}$ are all constant, so
that the only internal DOFs are those associated with $\{\theta_k\}$. Define
$\avecu{h}_k$ to be the rotation axis of the joint between bonds $k-1$ and $k$ that
is fixed in the frame of bond $k-1$; in the present case $\avecu{h}_k \equiv
\avecu{z}_{k-1}$. Insofar as indexing is concerned, there are $n_r$ internal
rotational joints (with labels $1, \ldots, n_r$), $n_b = n_r+1$ bonds ($0, \ldots,
n_r$) and $n_r+2$ sites ($0, \ldots, n_r+1$). In order to completely specify the
chain configuration, an additional joint is attached to the $k=0$ site, with three
translational and three rotational DOFs (conceptually equivalent to a telescopic
ball-and-socket joint); this joint is included in the formalism but will,
eventually, be treated separately.

\subsection{Kinematic and dynamic relations}

If $\avec{v}_k$ and $\avec{\omega}_k$ are the linear and angular velocities of site
$k$, then the velocities and accelerations of adjacent sites are related by
\begin{eqnarray}
\avec{\omega}_k &=& \avec{\omega}_{k-1} + \avecu{h}_k \dot\theta_k 
\label{eq:vdef} \\
\avec{v}_k &=& \avec{v}_{k-1} + \avec{\omega}_{k-1} \times \avec{b}_{k-1}
\label{eq:wdef} \\
\dot\avec{\omega}_k &=& \dot\avec{\omega}_{k-1} + \avecu{h}_k \ddot\theta_k +
\avec{\omega}_{k-1} \times \avecu{h}_k \dot\theta_k \label{eq:dvdef} \\
\dot\avec{v}_k &=& \dot\avec{v}_{k-1} + \dot\avec{\omega}_{k-1} \times
\avec{b}_{k-1} \nonumber \\
&& + \avec{\omega}_{k-1} \times (\avec{\omega}_{k-1} \times
\avec{b}_{k-1}) \label{eq:dwdef}
\end{eqnarray}
where $1 \le k \le n_r$. While the mass elements of the chain are normally
identified with the sites, here it is helpful to associate them with the bonds; if
$\avec{r}_k + \avec{c}_k$ is the location of the center-of-mass of the atoms
attached to bond $k$, then the center-of-mass acceleration of the bond is
\begin{equation}
\dot{\avec{v}^c_k} = \dot\avec{v}_k + \dot\avec{\omega}_k \times \avec{c}_k +
\avec{\omega}_k \times (\avec{\omega}_k \times \avec{c}_k) . \label{eq:dvcm}
\end{equation}
If $\avec{f}_k$ and $\avec{n}_k$ are the force and torque acting on bond $k$ across
joint $k$, then the equations of motion are
\begin{eqnarray}
\bm{\mathcal{I}}_k \dot\avec{\omega}_k + \avec{\omega}_k \times (\bm{\mathcal{I}}_k
\avec{\omega}_k) &=& \avec{n}_k - \avec{n}_{k+1} -
\avec{c}_k \times \avec{f}_k \nonumber \\
&& - (\avec{b}_k - \avec{c}_k) \times \avec{f}_{k+1} + \avec{n}^e_k
\label{eq:eqmang} \\
m_k \dot{\avec{v}^c_k} &=& \avec{f}_k - \avec{f}_{k+1} + \avec{f}^e_k ,
\label{eq:eqmlin}
\end{eqnarray}
where $\avec{f}^e_k$ and $\avec{n}^e_k$ are the externally applied force and
torque; $m_k$ and $\bm{\mathcal{I}}_k$ are the mass and moment of inertia of (the
atoms associated with) the bond, the latter expressed in a space-fixed frame and
relative to the center of mass of the bond. It is often convenient when dealing
with rigid bodies to work in a center-of-mass frame \cite{gol80}; this is not the
case here, and all vector components are expressed in the space-fixed coordinate
frame. Rearrange the terms of Eqs.~(\ref{eq:eqmang}) and (\ref{eq:eqmlin}) to
obtain relations between torques and forces on adjacent bonds,
\begin{eqnarray}
\avec{n}_k &=& \avec{n}_{k+1} + \avec{b}_k \times \avec{f}_{k+1} +
m_k \avec{c}_k \times \dot{\avec{v}^c_k} \nonumber \\
&& + \bm{\mathcal{I}}_k \dot\avec{\omega}_k +
\avec{\omega}_k \times (\bm{\mathcal{I}}_k \avec{\omega}_k) - \avec{n}^e_k -
\avec{c}_k \times \avec{f}^e_k \label{eq:ndef} \\
\avec{f}_k &=& \avec{f}_{k+1} + m_k \dot{\avec{v}^c_k} - \avec{f}^e_k ,
\label{eq:fdef}
\end{eqnarray}
and define the torque
\begin{equation}
t_k = \avecu{h}_k \cdot \avec{n}_k \label{eq:tdef}
\end{equation}
that acts along the axis $\avecu{h}_k$ at joint $k$ and corresponds to the
torsional interaction due to a twist around bond $k-1$.

\subsection{Spatial operator formulation}

Equations~(\ref{eq:vdef})--(\ref{eq:dwdef}) can be expressed more concisely in
terms of 6-component ``spatial'' vectors that combine the translational and
rotational quantities. It is also convenient to represent certain vectors by means
of antisymmetric matrices of form
\begin{equation}
\avect{u} = \left(\begin{array}{ccc} 0 & - u_z & u_y \\ u_z & 0 & - u_x \\
- u_y & u_x & 0 \end{array}\right) , 
\end{equation}
so that $\avect{u} \avec{v} \equiv \avec{u} \times \avec{v}$. The resulting
equations are
\begin{eqnarray}
\left(\begin{array}{c} \avec{\omega}_k \\ \avec{v}_k \end{array}\right) &=&
\left(\begin{array}{cc} \spa{I} & 0 \\ - \avect{b}_{k-1} & \spa{I}
\end{array}\right)
\left(\begin{array}{c} \avec{\omega}_{k-1} \\ \avec{v}_{k-1} \end{array}\right) +
\left(\begin{array}{c} \avecu{h}_k \\ 0 \end{array}\right) \dot\theta_k \\
\left(\begin{array}{c} \dot\avec{\omega}_k \\ \dot\avec{v}_k \end{array}\right)
&=& \left(\begin{array}{cc} \spa{I} & 0 \\ - \avect{b}_{k-1} & \spa{I} 
\end{array}\right)
\left(\begin{array}{c} \dot\avec{\omega}_{k-1} \\ \dot\avec{v}_{k-1} 
\end{array}\right) +
\left(\begin{array}{c} \avecu{h}_k \\ 0 \end{array}\right) \ddot\theta_k
\nonumber \\
&& + 
\left(\begin{array}{c} \avec{\omega}_{k-1} \times \avecu{h}_k \dot\theta_k \\
\avec{\omega}_{k-1} \times (\avec{\omega}_{k-1} \times \avec{b}_{k-1})
\end{array}\right)
\end{eqnarray}
or, equivalently,
\begin{eqnarray}
\spa{V}_k &=& \tr{\phi}_{k-1,k} \spa{V}_{k-1} + \tr{\spa{H}}_k \dot \spa{W}_k
\label{eq:vkdef} \\
\spa{A}_k &=& \tr{\phi}_{k-1,k} \spa{A}_{k-1} + \tr{\spa{H}}_k \ddot \spa{W}_k +
\spa{X}_k , \label{eq:akdef}
\end{eqnarray}
where $\spa{V}_k$ and $\spa{A}_k$ are examples of spatial vectors, and
\begin{equation}
\tr{\phi}_{k-1,k} = 
\left(\begin{array}{cc} \spa{I} & 0 \\ - \avect{b}_{k-1} & \spa{I}
\end{array}\right) .
\end{equation}
The $6 \times 6$ matrices $\tr{\phi}_{k-1,k}$ and $\phi_{k,k+1}$ (later) appear
throughout the derivation, and their role is to propagate kinematic and dynamic
information between joints. Several other new variables have been used:
\begin{equation}
\tr{\spa{H}}_k = \left(\begin{array}{c} \avecu{h}_k \\ 0 \end{array}\right) 
\end{equation}
is a 6-component joint axis vector (in the more general case of a joint with $d$
DOFs, which the formalism is capable of handling, $\tr{\spa{H}}_k$ would become a
$6 \times d$ matrix),
\begin{equation}
\spa{X}_k = \left(\begin{array}{cc} \avect{\omega}_{k-1} & 0 \\
0 & \avect{\omega}_{k-1} \end{array}\right) 
\left(\begin{array}{c} \avecu{h}_k \dot\theta_k \\
\avec{v}_k - \avec{v}_{k-1} \end{array}\right)
\end{equation}
is a 6-component spatial vector containing the remaining acceleration terms of the
current site, and $\dot\spa{W}_k \equiv \dot\theta_k$. When used in vectors and
matrices, $\spa{I}$ and $0$ denote unit and zero block submatrices of the implied
size. The 6-component vectors, and most of the associated matrices, are shown in
block capitals (to retain some similarity with \cite{jai93}, $\phi$, $\psi$ and
$\mathcal{M}$ are also used); no other special notation is needed since the
variable types will be obvious from the context.

In a similar way, Eqs.~(\ref{eq:ndef})--(\ref{eq:tdef}) can be rewritten as
\begin{eqnarray}
\left(\begin{array}{c} \avec{n}_k \\ \avec{f}_k \end{array}\right) &=&
\left(\begin{array}{cc} \spa{I} & \avect{b}_k \\ 0 & \spa{I} \end{array}\right)
\left(\begin{array}{c} \avec{n}_{k+1} \\ \avec{f}_{k+1} \end{array}\right) 
\nonumber \\
&& + \left(\begin{array}{c} m_k \avec{c}_k \times \dot{\avec{v}^c_k} +
\bm{\mathcal{I}}_k \dot\avec{\omega}_k + \avec{\omega}_k \times (\bm{\mathcal{I}}_k
\avec{\omega}_k) \\ m_k \dot{\avec{v}^c_k} \end{array}\right) \nonumber \\
&& - \left(\begin{array}{c} \avec{n}^e_k + \avec{c}_k \times \avec{f}^e_k \\
\avec{f}^e_k \end{array}\right) \\
\left(\begin{array}{c} t_k \\ 0 \end{array}\right) &=& 
\tr{\left(\begin{array}{c} \avecu{h}_k \\ 0 \end{array}\right)}
\left(\begin{array}{c} \avec{n}_k \\ \avec{f}_k \end{array}\right)
\end{eqnarray}
or, equivalently,
\begin{eqnarray}
\spa{F}_k &=& \phi_{k,k+1} \spa{F}_{k+1} + \spa{M}_k \spa{A}_k + \spa{Y}_k
\label{eq:fkdef} \\
\spa{T}_k &=& \spa{H}_k \spa{F}_k . \label{eq:tkdef}
\end{eqnarray}
Here Eq.~(\ref{eq:dvcm}) has been used, and
\begin{equation}
\spa{M}_k = \left(\begin{array}{cc} \bm{\mathcal{I}}_k - m_k \avect{c}_k
\avect{c}_k & m_k \avect{c}_k \\
- m_k \avect{c}_k & m_k \spa{I} \end{array}\right) , \label{eq:mkdef}
\end{equation}
is the symmetric, $6 \times 6$ mass matrix; the 6-component vector
\begin{equation}
\spa{Y}_k = 
\left(\begin{array}{c} \avect{\omega}_k (\bm{\mathcal{I}}_k - m_k \avect{c}_k
\avect{c}_k)
\avec{\omega}_k \\
m_k \avect{\omega}_k \avect{\omega}_k \avec{c}_k \end{array}\right) -
\left(\begin{array}{c} \avec{n}^e_k + \avec{c}_k \times \avec{f}^e_k \\
\avec{f}^e_k \end{array}\right) \label{eq:ykdef}
\end{equation}
contains the remaining force contributions. The identity $\avec{c}_k \times
(\avec{\omega}_k \times (\avec{\omega}_k \times \avec{c}_k)) = - \avec{\omega}_k
\times (\avec{c}_k \times (\avec{c}_k \times \avec{\omega}_k))$ was used in
obtaining these expressions. In order to use the recurrence relations for
$\spa{V}_k$, $\spa{A}_k$ and $\spa{F}_k$, the velocity and acceleration of the
initial site, $\spa{V}_0$ and $\spa{A}_0$, must be provided, while the force
associated with the site at the end of the final bond, $\spa{F}_{n_r+1}$, is zero,
since there is no joint associated with that site.

The purpose of the recurrence relations in Eqs.~(\ref{eq:akdef}) and
(\ref{eq:fkdef}) is to provide expressions for $\{\ddot \spa{W}_k\}$, which,
together with $\spa{A}_0$, and assuming all the forces acting on the sites are
known, can be integrated to solve for the chain dynamics; this is actually the
opposite of the typical robotics problem, in which the goal is to determine the
forces required to produce a particular robot arm trajectory.

\subsection{Stacked operators}

Equations~(\ref{eq:vkdef}), (\ref{eq:akdef}), (\ref{eq:fkdef}) and (\ref{eq:tkdef})
can be rewritten in condensed, ``stacked'' form
\begin{eqnarray}
\spa{V} &=& \tr{\phi} \spa{V} + \tr{\spa{H}} \dot \spa{W} \label{eq:vspacep} \\
\spa{A} &=& \tr{\phi} \spa{A} + \tr{\spa{H}} \ddot \spa{W} + \spa{X}
\label{eq:aspacep} \\
\spa{F} &=& \phi \spa{F} + \spa{M} \spa{A} + \spa{Y} \label{eq:fspacep} \\
\spa{T} &=& \spa{H} \spa{F} \label{eq:tspacep}
\end{eqnarray}
that combines the entire set of $k$ values. A quantity such as $\spa{V}$ containing
all the $\spa{V}_k$ values for the chain is also referred to as a spatial vector,
while, for example, the block matrix $\phi$ containing all the $\phi_{k,k+1}$
matrices is a spatial operator. The stacked formalism leads to a concise and
elegant formulation of the problem, free from inundation by indices as is often the
case in the robotics literature, e.g., \cite{fu87}.

The spatial operator approach was originally developed for the case of a fixed
initial bond \cite{rod92} -- the base in the example of a robot arm -- for which
$\spa{V}_0 = 0$, so that $\dot \spa{W} = {\rm col} (\dot\theta_1, \ldots,
\dot\theta_{n_r})$ is a vector with just $n_r$ components, and the other vectors
and matrices are sized accordingly. In order to remove the fixed-base restriction
\cite{jai95}, six extra DOFs are added to the problem by redefining $\dot \spa{W} =
{\rm col} ({\spa{V}}_0, \dot\theta_1, \ldots, \dot\theta_{n_r})$ as a vector with
$n_r+6$ components; likewise for $\ddot \spa{W}$. The size of the original  $6 n_r
\times n_r$ block-diagonal matrix $\spa{H} = {\rm diag} (\spa{H}_1, \ldots,
{\spa{H}}_{n_r})$ is increased to $6(n_r+1) \times (n_r+6)$ by including an extra
$6 \times 6$ block $\spa{H}_0 = \spa{I}$, so that now $\spa{H} = {\rm diag}
(\spa{I}, \spa{H}_1, \ldots, \spa{H}_{n_r})$. The block-diagonal matrix $\spa{M}$
is of size $6(n_r+1) \times 6(n_r+1)$; $\phi$ has the same size, and its only
non-zero blocks are those to the immediate right of the diagonal, namely
$\{\phi_{01}, \ldots, \phi_{n_r-1,n_r}\}$. Vectors $\spa{V}$, $\spa{A}$, $\spa{F}$,
$\spa{X}$, and $\spa{Y}$, all have $6(n_r+1)$ components, e.g., $\spa{V} = {\rm
col} (\spa{V}_0, \ldots, \spa{V}_{n_r})$, and $\spa{T}$ is organized in the same
way as $\dot {\spa{W}}$, with $n_r+6$ components; $\spa{T}_0 = 0$ because the
special $k=0$ joint exerts no torque. (Note that index order has been reversed from
the original to make it more suitable for polymer use, and, for convenience, other
aspects of the notation have been altered or simplified.)

The next step is to define the matrix
\begin{equation}
\Phi = (\spa{I} - \phi)^{-1} , \label{eq:phidef}
\end{equation}
which is also used in the alternative form, $\Phi = \Phi \phi + \spa{I}$; because
$\phi^{n_r + 1} = 0$, Eq.~(\ref{eq:phidef}) is equivalent to $\Phi = \spa{I} + \phi
+ \phi^2 + \cdots + \phi^{n_r}$, which is an upper-triangular block matrix whose
elements, each a $6 \times 6$ matrix, are
\begin{equation}
\Phi_{ij} = \left\{ \begin{array}{ll} \spa{I} & \quad j = i \\
\phi_{i,i+1} & \quad j = i + 1 \\
\phi_{i,i+1} \cdots \phi_{j-1,j} & \quad j > i + 1 . \end{array}\right.
\end{equation}
Then, in terms of $\Phi$, Eqs.~(\ref{eq:vspacep})--(\ref{eq:tspacep}) reduce to
\begin{eqnarray}
\spa{V} &=& \tr{\Phi} \tr{\spa{H}} \dot \spa{W} \\
\spa{A} &=& \tr{\Phi} (\tr{\spa{H}} \ddot \spa{W} + \spa{X}) \label{eq:addw} \\
\spa{T} &=& \mathcal{M} \ddot \spa{W} + \spa{H} \Phi (\spa{M} \tr{\Phi}
\spa{X} + \spa{Y}) , \label{eq:tmddw}
\end{eqnarray}
where
\begin{equation}
\mathcal{M} = \spa{H} \Phi \spa{M} \tr{\Phi} \tr{\spa{H}} . \label{eq:mmdef}
\end{equation}

While $\spa{M}$ is a sparse, $6(n_r+1) \times 6(n_r+1)$ block-diagonal matrix,
${\cal M}$ is only of size $(n_r+6) \times (n_r+6)$, but, although it is typically
much smaller, it is fully populated. In principle Eq.~(\ref{eq:tmddw}) can be
numerically integrated to obtain $\spa{W}$, and this is one of the approaches
actually used in solving the problem, but the computational effort required for
evaluating $\mathcal{M}^{-1}$ at each timestep to obtain $\ddot \spa{W}$ is of
order $O\big((n_r+6)^3\big)$; for this reason such an approach is not practical for
any but the shortest of chains. The alternative method, described below, requires a
computational effort of order $O(n_r)$, together with what amounts to the inversion
of a $6 \times 6$ matrix; clearly this will prove to be a far more efficient
approach, even for relatively small $n_r$.

\subsection{Inversion of the mass matrix}

As a preliminary step in obtaining an explicit expression for $\mathcal{M}^{-1}$
define \cite{rod92} the $6 \times 6$ matrix $\spa{P}_k$ in terms of $\spa{M}_k$ as
\begin{equation}
\spa{P}_k = \phi_{k,k+1} (\spa{I} - \spa{G}_{k+1} \spa{H}_{k+1}) \spa{P}_{k+1}
\tr{\phi}_{k,k+1} + \spa{M}_k . \label{eq:pkdef}
\end{equation}
In Eq.~(\ref{eq:pkdef}),
\begin{eqnarray}
\spa{G}_k &=& \spa{P}_k \tr{\spa{H}}_k \spa{D}^{-1}_k \label{eq:gkdef} \\
\spa{D}_k &=& \spa{H}_k \spa{P}_k \tr{\spa{H}}_k , \label{eq:dkdef}
\end{eqnarray}
where, for joints with a single DOF, $\spa{G}_k$ is a 6-component vector and
$\spa{D}_k$ is a nonzero scalar; note also that $\spa{P}_k$ is symmetric. (The
motivation for introducing $\spa{P}_k$ is explained in \cite{rod92} and derives
from the formal similarity of these equations with those used in the completely
unrelated field of linear filtering.) Also define
\begin{equation}
\psi_{k,k+1} = \phi_{k,k+1} (\spa{I} - \spa{G}_{k+1} \spa{H}_{k+1})
\label{eq:psidef}
\end{equation}
and substitute this in Eq.~(\ref{eq:pkdef}). The stacked versions of
Eqs.~(\ref{eq:pkdef})--(\ref{eq:psidef}) are
\begin{eqnarray}
\spa{P} &=& \psi \spa{P} \tr{\phi} + \spa{M} \label{eq:spdef} \\
\spa{G} &=& \spa{P} \tr{\spa{H}} \spa{D}^{-1} \label{eq:sgdef} \\
\spa{D} &=& \spa{H} \spa{P} \tr{\spa{H}} \label{eq:sddef} \\
\psi &=& \phi (\spa{I} - \spa{G} \spa{H}) . \label{eq:spsidef}
\end{eqnarray}
Matrices $\spa{P}$ and $\psi$ are of size $6(n_r+1) \times 6(n_r+1)$, and $\spa{G}$
is $(n_r+6) \times 6(n_r+1)$ and block-diagonal (thus the product $\spa{G}_{k+1}
\spa{H}_{k+1}$ is square). Matrix $\spa{D}$ is of size $(n_r+6) \times (n_r+6)$;
its first $6 \times 6$ diagonal block corresponds to $\spa{D}_0$, and the remaining
$n_r$ diagonal elements are the scalars $\spa{D}_k$. From Eqs.~(\ref{eq:spdef}) and
(\ref{eq:spsidef}),
\begin{equation}
\spa{M} = \spa{P} - \phi \spa{P} \tr{\phi} + \phi \spa{G} \spa{H} \spa{P}
\tr{\phi} , \label{eq:mdef}
\end{equation}
and so, by using Eq.~(\ref{eq:phidef}),
\begin{equation}
\Phi \spa{M} \tr{\Phi} = \spa{P} + \Phi \phi \spa{P} + \spa{P} \tr{\phi} \tr{\Phi} +
\Phi \phi \spa{P} \tr{\spa{H}} \spa{D}^{-1} \spa{H} \spa{P} \tr{\phi} \tr{\Phi} .
\label{eq:pmp} 
\end{equation}
Substitute Eq.~(\ref{eq:pmp}) in (\ref{eq:mmdef}), then use $\spa{G} \spa{D} =
{\spa{P}} \tr{\spa{H}}$ from (\ref{eq:sgdef}), together with (\ref{eq:sddef}), to
obtain
\begin{eqnarray}
\mathcal{M} &=& \spa{H} \spa{P} \tr{\spa{H}} + \spa{H} \Phi \phi \spa{P}
\tr{\spa{H}} + \spa{H} \spa{P} \tr{\phi} \tr{\Phi} \tr{\spa{H}} \nonumber \\
&& + \spa{H} \Phi \phi \spa{P} \tr{\spa{H}} \spa{D}^{-1} \spa{H} \spa{P}
\tr{\phi} \tr{\Phi} \tr{\spa{H}} \nonumber \\
&=& (\spa{I} + \spa{H} \Phi \phi \spa{G}) \spa{D} \tr{(\spa{I} +
\spa{H} \Phi \phi \spa{G})} . \label{eq:mfac}
\end{eqnarray}
This alternative factorization of $\mathcal{M}$ is a product of three $(n_r+6)
\times (n_r+6)$ matrices, unlike Eq.~(\ref{eq:mmdef}) which involves nonsquare
matrices.

It is now a straightforward matter to invert $\mathcal{M}$. Use a special case of
the Woodbury formula for the inverse of a matrix \cite{pre92}, $(\spa{I} +
\spa{Q}_1 \spa{Q}_2)^{-1} = \spa{I} - \spa{Q}_1 (\spa{I} + \spa{Q}_2
\spa{Q}_1)^{-1} \spa{Q}_2$ to write
\begin{equation}
(\spa{I} + \spa{H} \Phi \phi \spa{G})^{-1} = \spa{I} - \spa{H} \Phi (\spa{I} +
\phi \spa{G} \spa{H} \Phi)^{-1} \phi \spa{G} . 
\end{equation}
By analogy with Eq.~(\ref{eq:phidef}) for $\Phi$, define $\Psi = (\spa{I} -
\psi)^{-1}$; then from Eqs.~(\ref{eq:spsidef}) and (\ref{eq:phidef}),
\begin{equation}
\Psi^{-1} = \Phi^{-1} + \phi \spa{G} \spa{H} , \label{eq:psiphi}
\end{equation}
so that $(\spa{I} + \spa{H} \Phi \phi \spa{G})^{-1} = \spa{I} - \spa{H} \Psi \phi
\spa{G}$. Thus the inverse of Eq.~(\ref{eq:mfac}) is
\begin{equation}
\mathcal{M}^{-1} = \tr{(\spa{I} - \spa{H} \Psi \phi \spa{G})} \spa{D}^{-1}
(\spa{I} - \spa{H} \Psi \phi \spa{G}) , 
\end{equation}
and so, from Eq.~(\ref{eq:tmddw}),
\begin{eqnarray}
\ddot \spa{W} &=& \tr{(\spa{I} - \spa{H} \Psi \phi \spa{G})} \spa{D}^{-1}
(\spa{I} - \spa{H} \Psi \phi \spa{G}) \nonumber \\
&& \times [\spa{T} - \spa{H} \Phi (\spa{M} \tr{\Phi} \spa{X} + \spa{Y})]
\nonumber \\
&=& \tr{(\spa{I} - \spa{H} \Psi \phi \spa{G})} \spa{D}^{-1} \nonumber \\
&& \times [\spa{T} - \spa{H} \Psi
(\phi \spa{G} \spa{T} + \spa{M} \tr{\Phi} \spa{X} + \spa{Y})] , \label{eq:ddw} 
\end{eqnarray}
where Eq.~(\ref{eq:psiphi}) is used in simplifying $\spa{H} (\spa{I} - \Psi \phi
\spa{G} \spa{H}) \Phi = \spa{H} \Psi$. To eliminate $\Psi$, first rewrite
Eq.~(\ref{eq:ddw}) as
\begin{equation}
\tr{(\spa{I} + \spa{H} \Phi \phi \spa{G})} \ddot \spa{W} = \spa{D}^{-1} [\spa{T} -
\spa{H} \Psi (\phi \spa{G} \spa{T} + \spa{M} \tr{\Phi} \spa{X} + \spa{Y})] .
\end{equation}
Next, use Eq.~(\ref{eq:spdef}) with (\ref{eq:phidef}) to get
\begin{eqnarray}
\Psi \spa{M} \tr{\Phi} &=& \Psi \spa{P} (\tr{\phi} \tr{\Phi} + \spa{I}) -
\Psi \psi \spa{P} \tr{\phi} \tr{\Phi} \nonumber \\
&=& \Psi \spa{P} + \spa{P} \tr{\phi} \tr{\Phi} .
\end{eqnarray}
Then, using the transpose of Eq.~(\ref{eq:sgdef}), it follows that
\begin{equation}
\tr{(\spa{I} + \spa{H} \Phi \phi \spa{G})} \ddot \spa{W} = \spa{D}^{-1} \spa{E} -
\tr{\spa{G}} \tr{\phi} \tr{\Phi} \spa{X} , \label{eq:ddwnext}
\end{equation}
in which the force-like quantities
\begin{eqnarray}
\spa{E} &=& \spa{T} - \spa{H} \spa{Z} \label{eq:epsdef} \\
\spa{Z} &=& \Psi (\phi \spa{G} \spa{T} + \spa{P} \spa{X} + \spa{Y}) 
\label{eq:zdeforig}
\end{eqnarray}
have been defined. Rearranging Eq.~(\ref{eq:ddwnext}) and using the expression for
$\spa{A}$ given in Eq.~(\ref{eq:addw}) leads to
\begin{eqnarray}
\ddot \spa{W} &=& \spa{D}^{-1} \spa{E} - \tr{\spa{G}} \tr{\phi} \tr{\Phi}
(\tr{\spa{H}} \ddot \spa{W} + \spa{X}) \nonumber \\
&=& \spa{D}^{-1} \spa{E} - \tr{\spa{G}} \tr{\phi} \spa{A} . \label{eq:ddwdef}
\end{eqnarray}
It is also possible to eliminate $\Psi$ from Eq.~(\ref{eq:zdeforig}) by
substituting $\spa{T}$ from (\ref{eq:epsdef}) to get $(\spa{I} - \Psi \phi \spa{G}
\spa{H}) \spa{Z} = \Psi (\phi \spa{G} \spa{E} + \spa{P} \spa{X} + \spa{Y})$, and
then using Eq.~(\ref{eq:psiphi}) to obtain
\begin{equation}
\spa{Z} = \Phi (\phi \spa{G} \spa{E} + \spa{P} \spa{X} + \spa{Y}) .
\label{eq:zdef}
\end{equation}

Explicit forms for the new recurrence relations embodied in Eqs.~(\ref{eq:ddwdef})
and (\ref{eq:zdef}) are obtained by using Eq.~(\ref{eq:phidef}) and reintroducing
the $k$ indices --
\begin{eqnarray}
\spa{Z}_k &=& \phi_{k,k+1} (\spa{Z}_{k+1} + \spa{G}_{k+1} \spa{E}_{k+1}) +
\spa{P}_k \spa{X}_k + \spa{Y}_k \label{eq:zkdef} \\
\ddot \spa{W}_k &=& \spa{D}^{-1}_k \spa{E}_k - \tr{\spa{G}}_k \tr{\phi}_{k-1,k}
\spa{A}_{k-1} \label{eq:ddsgimakdef} .
\end{eqnarray}
These recurrence relations are used in opposite $k$-directions; they succeed in 
providing the required results without the need for explicit evaluation of the
matrix inverse $\mathcal{M}^{-1}$ as implied by Eq.~(\ref{eq:tmddw}). It is for
this reason that the method has not been referred to as an ``inverse matrix
method'', a term sometimes seen in the literature, but rather a ``rigid link''
method, a far more apt descriptor.

The expressions given here describe the entire chain, but, provided the end joints
are handled correctly, these results can be used for linear segments that form part
of a larger assembly, allowing more complicated tree-like structures to be treated.
Furthermore, while the above formulation deals with the simplest case of a linear
chain with a single torsional DOF per joint, it is readily extended to more complex
joints, enabling, for example, the constant bond-angle condition to be eliminated
by allowing two DOFs at each joint (an alternative would be to decompose an
individual joint into two coincident joints each with a single DOF).

\section{Simulation techniques}

\subsection{Linked-chain equations of motion}

The recurrence relations used to propagate velocities, forces, and accelerations
along the chain are as follows: The (translational and rotational) velocities
$\spa{V}_k$ are obtained by starting with $\spa{V}_0$ and iterating
Eq.~(\ref{eq:vkdef}),
\begin{equation}
\spa{V}_k = \tr{\phi}_{k-1,k} \spa{V}_{k-1} + \tr{\spa{H}}_k \dot \spa{W}_k,
\quad k = 1, \ldots, n_r . \label{eq:recurv}
\end{equation}
The forces (and torques), as represented by $\spa{E}_k$, together with the matrices
$\spa{D}_k$ and $\spa{G}_k$, are obtained by iterating Eqs.~(\ref{eq:pkdef}) and
(\ref{eq:zkdef}). For computational convenience, new quantities $\spa{A}'_k$ and
$\spa{Z}'_k$ are introduced; then, starting with $\spa{P}_{n_r+1} = 0$ and
$\spa{Z}'_{n_r+1} = 0$,
\begin{equation}
\left.\begin{array}{ccl}
\spa{P}_k &=& \phi_{k,k+1} (\spa{I} - \spa{G}_{k+1} \spa{H}_{k+1}) \nonumber \\
&& \times \spa{P}_{k+1} \tr{\phi}_{k,k+1} + \spa{M}_k \\
\spa{D}_k &=& \spa{H}_k \spa{P}_k \tr{\spa{H}}_k \\
\spa{G}_k &=& \spa{P}_k \tr{\spa{H}}_k \spa{D}^{-1}_k \\
\spa{Z}_k &=& \phi_{k,k+1} \spa{Z}'_{k+1} + \spa{P}_k \spa{X}_k + \spa{Y}_k \\
\spa{E}_k &=& \spa{T}_k - \spa{H}_k \spa{Z}_k \\
\spa{Z}'_k &=& \spa{Z}_k + \spa{G}_k \spa{E}_k
\end{array} \quad\right\} 
\quad k = n_r, \ldots, 0 . \label{eq:recurpz}
\end{equation}
Finally, the values of $\ddot \spa{W}_k$ (or $\ddot\theta_k$) are determined by
starting with $\spa{A}_0$ (its evaluation is discussed below), and iterating
Eqs.~(\ref{eq:akdef}) and (\ref{eq:ddsgimakdef}),
\begin{equation}
\left.\begin{array}{ccl}
\spa{A}'_k &=& \tr{\phi}_{k-1,k} \spa{A}_{k-1} \\
\ddot \spa{W}_k &=& \spa{D}^{-1}_k \spa{E}_k - \tr{\spa{G}}_k \spa{A}'_k \\
\spa{A}_k &=& \spa{A}'_k + \tr{\spa{H}}_k \ddot \spa{W}_k + \spa{X}_k \\
\end{array} \quad\right\}
\quad k = 1, \ldots, n_r . \label{eq:recura}
\end{equation}
These recurrence relations, which are readily transformed into a suitable computer
program, imply a series of operations (multiplications and additions) involving $6
\times 6$ matrices and 6-component vectors, but the total computational effort is
only of order $O(n_r)$.
 
Recall that the $k=0$ joint has six DOFs, and also that $\spa{H}_0 = \spa{I}$,
$\spa{X}_0 = 0$, and $\ddot \spa{W}_0 = \spa{A}_0$. Now, because $\spa{A}_{-1} =
0$, it follows from Eq.~(\ref{eq:recura}) that $\spa{A}_0 = \spa{D}^{-1}_0
\spa{E}_0$, and since $\spa{T}_0 = 0$,
\begin{equation}
\spa{D}_0 \spa{A}_0 = - \spa{Z}_0 , \label{eq:dzero}
\end{equation}
where both $\spa{D}_0$ and $\spa{Z}_0$ have already been determined (above). Thus
$\spa{A}_0$ can be evaluated numerically by solving the set of six linear equations
contained in Eq.~(\ref{eq:dzero}) using the standard LU method \cite{pre92}; the
computational effort required for this initial joint is fixed and independent of
$n_r$.

\subsection{Leapfrog integration and rigid-body equations}

The familiar leapfrog method for integrating the MD translational equations of
motion -- which is algebraically equivalent to the Verlet method \cite{ver67} -- is
usually expressed in a form where the coordinates and velocities are evaluated at
alternate half-timesteps \cite{rap95}. This minor inconvenience can be avoided by
using a slightly modified form that breaks the integration procedure for a single
timestep into two parts: Prior to computing the latest acceleration ($\avec{a}$)
values, update the velocities ($\avec{v}$) by a half timestep using the previous
accelerations, and then update the coordinates ($\avec{r}$) by a full timestep
using these intermediate velocity values,
\begin{eqnarray}
\avec{v}(t + h / 2) &=& \avec{v}(t) + (h / 2) \avec{a}(t) \\
\avec{r}(t + h) &=& \avec{r}(t) + h \avec{v}(t + h / 2) .
\end{eqnarray}
In the case of the polymer chain, this procedure is applied to the translation
coordinates of the $k=0$ site and (in scalar form) to each of the dihedral angles
$\theta_k$; the treatment of the angular coordinates associated with the $k=0$
site, below, employs a related approach for dealing with the rotational equations.
Next, use the new coordinates (and velocities if needed) to compute the latest
acceleration values, then update the velocities over the second half timestep,
\begin{equation}
\avec{v}(t + h) = \avec{v}(t + h / 2) + (h / 2) \avec{a}(t + h) . 
\end{equation}

In the linked-chain formulation, the initial bond of the chain is treated as a
rigid body; the influence of the rest of the chain on it has already been taken
into account and is contained in the force and torque transmitted through the first
internal joint. There are a number of ways of describing the orientation of a rigid
body \cite{gol80}: Euler angles have proved very useful for analytic purposes
because of their intuitive nature, but owing to a potentially singular matrix that
appears in the equations of motion they are not the preferred method for dealing
with numerical problems. Quaternions have achieved popularity because of their
singularity-free nature, but their normalization must be preserved against a small
but persistent numerical drift \cite{eva77,rap95}. A more recently proposed
alternative is to regard the complete rotation matrix as the dynamical variable;
this is the representation that will be used here, since the integration scheme
\cite{dul97} -- which is based on operator splitting and maintains
time-reversibility -- is just another instance of the leapfrog method.

In the original description \cite{dul97}, vector components were expressed in the
principal-axis frame of the body. Since the chain dynamical problem as a whole is
solved in the space-fixed frame, the corresponding form of the rotational equations
will be described here. If $\amat{R}$ denotes the rotation matrix of a rigid
body, then the first part of the leapfrog integration step consists of a
half-timestep update of the angular velocities,
\begin{equation}
\avec\omega(t + h / 2) = \avec\omega(t) + (h / 2) \avec\alpha(t) , 
\end{equation}
where $\avec\alpha \equiv \dot\avec\omega$, followed by a full-timestep update of
$\amat{R}$ using a symmetric product of matrices describing a series of small
partial rotations,
\begin{equation}
\tr{\amat{R}}(t + h) = \amat{U}_1 \, \amat{U}_2 \, \amat{U}_3 \, \amat{U}_2 \,
\amat{U}_1 \, \tr{\amat{R}}(t) , 
\end{equation}
where, for convenience, the transpose of $\amat{R}$ is treated. Note that for the
linked chain, the rigid body is associated with the $k=0$ joint, so that $\amat{R}
\equiv \amat{R}_0$. Each of the matrices
\begin{equation}
\amat{U}_1 = \amat{U}_x (\omega_x h / 2),
\ \amat{U}_2 = \amat{U}_y (\omega_y h / 2),
\ \amat{U}_3 = \amat{U}_z (\omega_z h) 
\end{equation}
describes a rotation about a single axis and is evaluated in the space-fixed frame.
For small angles, they can be approximated in a way that preserves orthogonality,
e.g.,
\begin{equation}
\amat{U}_x(\theta) = 
\left(\begin{array}{ccc} 1 & 0 & 0 \\
0 & \cos \theta & - \sin \theta \\ 
0 & \sin \theta & \hphantom{-} \cos \theta \end{array}\right)
\approx \left(\begin{array}{ccc} 1 & 0 & 0 \\
0 & {1 - \theta^2 / 4 \over 1 + \theta^2 / 4}
& {- \theta \over 1 + \theta^2 / 4} \\
0 & {\theta \over 1 + \theta^2 / 4}
& {1 - \theta^2 / 4 \over 1 + \theta^2 / 4} \end{array}\right) .
\end{equation}
The second part of the leapfrog step is
\begin{equation}
\avec\omega(t + h) = \avec\omega(t + h / 2) + (h / 2) \avec\alpha(t + h) .
\end{equation}
In the case of a single rigid body, the angular acceleration is determined from the
torque $\avec\tau$, namely $\avec\alpha(t + h) = {\bm{\mathcal{I}}}^{-1} \,
\avec\tau(t + h)$, whereas for the linked chain this treatment is only required for
the $k=0$ joint, and $\avec\alpha$ is obtained by solving Eq.~(\ref{eq:dzero}); the
reason rigid bodies are usually treated in the body-fixed principal-axes frame is
to ensure the diagonality of $\bm{\mathcal{I}}$, a consideration that is not
relevant here.

The complete procedure for a single timestep can be summarized as the following
sequence of operations: integrate (first part) to obtain base velocities and
coordinates, and joint angular velocities and angles; determine site velocities,
Eq.~(\ref{eq:recurv}); evaluate site coordinates,
Eqs.~(\ref{eq:rotmat})--(\ref{eq:sitecoord}); compute external forces and torques,
and other necessary quantities; determine joint forces, Eq.~(\ref{eq:recurpz});
solve Eq.~(\ref{eq:dzero}) for the base acceleration; determine joint
accelerations, Eq.~(\ref{eq:recura}); integrate (second part) to obtain base
velocities and joint angular velocities.

\subsection{Polymer chain model}

Two kinds of interactions are required in this model -- excluded volume and
torsion. The former is provided by a pair interaction that prevents overlap of the
atoms (or atom groups) located at the chain sites. Here a simple soft-sphere
repulsion, based on the Lennard-Jones potential with a short-range cutoff, is all
that is required: for a pair of atoms located at $\avec{r}_i$ and $\avec{r}_j$,
where $\avec{r}_{ij} = \avec{r}_i - \avec{r}_j$, and $r_{ij} = | \avec{r}_{ij}|$,
the potential is
\begin{equation}
u_{ss}(r_{ij}) = \left\{\begin{array}{cc}
4 \epsilon [(r_{ij} / \sigma)^{12} - (r_{ij} / \sigma)^6] & \quad r_{ij} < r_c \\
0 \hfill & \quad r_{ij} \ge r_c \end{array}\right. \label{eq:sspot}
\end{equation}
with a cutoff $r_c = 2^{1/6} \sigma$ (nearby pairs of atoms that are prevented from
approaching too closely because of geometrical restrictions need not be
considered). Should a pairwise attraction between particular pairs of distant chain
atoms be required (as will be the case later on), it can be obtained from
Eq.~(\ref{eq:sspot}) by simply increasing $r_c$. The pair forces derived from this
potential, and their associated torques, contribute to $\avec{f}^e_k$ and
$\avec{n}^e_k$ in Eqs.~(\ref{eq:eqmang}) and (\ref{eq:eqmlin}).

The torsional potential associated with the dihedral angles $\theta_k$ has the
simple form
\begin{equation}
u_t(\theta_k) = - u_k \cos (\theta_k - \theta_k^{(0)}) , \label{eq:torspot}
\end{equation}
where $\theta_k^{(0)}$ is the dihedral angle that produces a ground state having
the correct helical twist, and $u_k$ is the interaction strength. The torque
appearing in Eq.~(\ref{eq:tdef}) is
\begin{equation}
t_k = u_k \sin (\theta_k - \theta_k^{(0)}) ,
\end{equation}
a result whose simplicity stands in sharp contrast to the intricate vector algebra
associated with torque calculations when working in Cartesian coordinates
\cite{rap95}.

For the chains considered here it is assumed all $|\avec{b}_k| = b$, $\alpha_k =
\alpha$, $\theta_k^{(0)} = \theta^{(0)}$ and, except for the later twin-helix
studies where selected $u_k = 0$, all $u_k = u^{(0)}$. Since the torsion also acts
at the first internal joint, it is necessary to add an extra site and bond to the
chain (effectively with an index ``$-1$'') to make this torsion term meaningful;
the first three sites of the modified chain form a rigid unit (the extra bond does
not alter the preceding analysis) and the chain length is increased by unity.

A spherical mass element (with a finite moment of inertia about its own center of
mass) is associated with each site; for bonds with $k>0$, the mass is attached to
the far ($k+1$ site) of the bond, while the $k=0$ bond, as explained above, has
three masses associated with it. The components of the inertia tensor in
Eqs.~(\ref{eq:mkdef}) and (\ref{eq:ykdef}) are
\begin{equation}
(\bm{\mathcal{I}}_k)_{ij} = \left\{ \begin{array}{ll}
\hphantom{-} \sum_{\kappa \in k} m_\kappa (r_\kappa^2 - r_{\kappa i}^2) &
\quad i = j \\
- \sum_{\kappa \in k} m_\kappa r_{\kappa i} r_{\kappa j} & \quad i \ne j
\end{array}\right. ,
\end{equation}
where the sum (or volume integral) is over all mass elements $\kappa$ fixed to bond
$k$, and coordinates are relative to the center of mass of each bond in the
space-fixed frame. 

\subsection{Order parameter}

While the appearance of an ordered helical structure, even one with the occasional
defect, is easily recognized visually, in order to facilitate statistical analysis
of the behavior it is important to be able to quantify the degree of order present
in the chain. Let
$\avec{d}_k = \avec{b}_{k-1} \times \avec{b}_k$, then 
\begin{equation}
S = {1 \over n_r} \sum_{k=1}^{n_r} \avecu{d}_k \label{eq:sdef}
\end{equation}
defines an order parameter that measures the long-range order present in the folded
structure based on the orientation of the helical turns; for a single, well-formed
helix, $S$ should have a value close to unity. A slightly modified version of $S$
will be introduced later for studying twin-helix structures.

This definition of $S$ is particularly useful for detecting structures consisting
of two or more helical domains with axes aligned in different directions due to a
localized defect of the type seen in helically wound telephone and electrical
cords. Since the correct helicity (or ``handedness'') is built into the
interactions, it is unlikely that segments of opposite helicity will independently
nucleate at separate locations but, as the chain collapses, individual turns with
the wrong twist can become trapped in the structure. These defects are capable of
traveling along the chain, but this is a slow process, and the direction of motion
is random unless close to the chain end. There are instances where the definition
of $S$ in Eq.~(\ref{eq:sdef}) can give an incomplete picture; if a wrong turn
occurs very close to the chain end, its effect on $S$ will be minimal, and even a
perfectly formed helix is subject to low frequency bending motion. Other order
parameters can be defined that are of a more short-range nature; for example, a
simple count of the number of pairs of chain sites lying within a specified range
(i.e., the number of ``contacts'' between adjacent turns of the helix) divided by
the maximum possible value, but for long chains the tolerance in the threshold
required to accommodate thermal fluctuations might allow significant changes in the
helical-axis direction to go undetected.

\section{Results}

\subsection{Simulation details}

The most prominently recognizable structural motif found in proteins is the
($\alpha$-) helix. A helix, because of its uniformity along the longitudinal axis,
is a particularly simple structure to specify, and both Monte Carlo and MD
helix-folding simulations based on the complex potentials designed for protein
modeling have been carried out, e.g., \cite{rap81,bvc98}. Since the complexity of
these potentials contributes little to a basic understanding of generic folding
phenomena, the present simulations are based on the much simpler model and
potentials described previously. Indeed, an analogous approach has been employed
experimentally \cite{nel97} in a study of helix formation in synthesized
nonbiological chain molecules, where the interactions are simpler than in proteins
(in particular, there are no hydrogen bonds).

The importance of examining simple structures, such as the helix, is that the
process by which ordered arrangements emerge from randomly coiled states is likely
to capture something of the essence of real protein folding, such as the
cooperativity of the folding process, the role of nucleation sites, the degree to
which folding is able to proceed to completion, and the steric and topological
effects of excluded volume. The key question, of course, is whether the intrinsic
timescales of these processes are sufficiently small for simulation to be
computationally feasible, an issue addressed by the results presented here. While
the present model is admittedly a mere caricature of the detailed models normally
employed in studies of individual proteins, it has two undeniable advantages,
namely a known native ground state compatible with the interactions, and
sufficiently modest computational requirements that MD simulation is able to
encompass the time interval required for major conformational change. More complex
protein structures also display certain common characteristics, and ought to be
accessible to simulations of this type; it is, however, essential to eliminate any
ambiguity from the ground state, something that nature itself has presumably
achieved in the interests of efficiency and reliability.

Each simulation run considers a single chain constructed as described earlier. The
absence of a solvent, apart from changing the timescales, should not alter the
outcome; indeed many, if not most, protein simulations avoid introducing an
explicit solvent for reasons of computational efficiency. The simulation is begun
at a relatively high temperature, so that the kinetic energy is sufficiently large
to surmount the torsional potential barriers. The initial chain configuration is a
large loop extending across the simulation cell, with a very slight helicity to
prevent any overlap; initial dihedral angles are chosen so that locally, the
conformation is almost a planar zigzag state. The joint angular velocities are
assigned random values corresponding to the starting temperature, and memory of
this initial state rapidly vanishes early in the simulation. The temperature is
gradually reduced by a factor slightly less than unity at regular intervals until,
towards the end of the run, very little kinetic energy remains in the system. The
simulation region is bounded by hard, reflecting walls; while there are occasional
wall collisions, this has little influence on the overall behavior. (The
alternative would be to use periodic boundaries, which for a simulation cell not
large enough to contain the chain in a fully-stretched state, would be subject to
chain wraparound effects; while these are also unlikely to affect the overall
behavior, they can prove visually confusing given the importance of
computer-generated visualization in this work.)

The gradual cooling that is imposed throughout the run plays several distinct
roles. During the early stage it is used to drive the chain from a totally random
state to one in which the torsional potential begins to have some influence over
the dihedral angles. Then, as the temperature is reduced further, an increasing
degree of local order emerges and precursors to long-range order appear, either as
a consequence of the merging of separate ordered domains, or the spread of order
from a nucleation region (or a combination of both processes); during this stage
the imposed cooling performs a task normally the responsibility of the solvent,
namely the removal of excess potential energy as the chain evolves towards states
of lower energy. Once the chain has reached a state consisting mainly of helical
segments, possibly separated by small misfolded regions that have become trapped,
the purpose of further temperature reduction is to gradually freeze out thermal
fluctuations -- without further major structural change -- in order to allow
evaluation of the long-range order parameter $S$ (the measure of success of the
folding process); the latter part of this cooling stage is not intended to imitate
any real physical process.

The simulations use standard, reduced MD units, in which all distances and energies
are expressed in terms of the Lennard-Jones parameters $\sigma$ and $\epsilon$
respectively; mass is expressed in terms of the monomer mass $m$ and, consequently,
the unit of time is $\sqrt{m \sigma^2 / \epsilon}$. Temperature and energy are made
numerically identical by setting the Boltzmann constant $k_{\rm B}$ to unity. In
terms of these units the parameters used in the runs are as follows: The bond
length $b = 1.3$, a value sufficiently short to prevent the chain crossing itself,
the bond angle $\alpha$ and the preferred dihedral angle $\theta^{(0)}$ are chosen
to produce helices with periodicity six, and the torsional potential strength
$u^{(0)} = 5$. In the studies of twin helices, the cutoff in the attractive
interaction, based on Eq.~(\ref{eq:sspot}), occurs at $r_c = 2.2$. The initial
temperature is 4 (corresponding to a kinetic energy per DOF of 2) and the final
temperature is $10^{-3}$; temperature is reduced by rescaling all velocities and
angular velocities by a factor $f_T$ every 4000 timesteps, with $f_T = 0.95$ or
0.97. The runs reported here are each of length 4--8 $\times 10^5$ steps; the
integration timestep (in MD units) is $h = 4 \times 10^{-3}$. In order to produce
reliable statistics, a large number of runs were carried out for each case studied;
the runs differed in the choice of initial random values for $\{\dot\theta_k\}$.

\subsection{Folding to a single helix }

Measurements were made of the long-range order parameter $S$ and the total energy,
the latter a sum over contributions from the soft-sphere pair interactions,
Eq.~(\ref{eq:sspot}), the torsional terms, Eq.~(\ref{eq:torspot}), and the kinetic
energy. The measurements involved 400 independent runs for each of several chain
lengths $L$ and different cooling rates. These quantitative results were
complemented by an interactive graphical version of the simulation program that
provided real-time visual monitoring of the folding process; in addition to
learning about any potential obstructions to complete folding, the ability to
observe chains directly also helped when choosing a cooling rate sufficiently fast
for folding to proceed to completion, but not too fast for an excessive number of
defects to become trapped in the nascent structures.

\begin{table}
\caption{\label{tab:runs}Details of helix folding runs discussed in the text.}
\begin{ruledtabular}
\begin{tabular}{ccccc}
Length ($L$) & Turns & $f_T$\footnote{Cooling factor.} & Steps ($\times 10^3$) &
Success\footnote{Criterion for successful helix formation is defined in the text.}
\\
\hline
18     &  3     &  0.95  &  400     &  1.00   \\  
36     &  6     &  0.95  &  400     &  0.94   \\
54     &  9     &  0.95  &  400     &  0.85   \\
54     &  9     &  0.97  &  800     &  0.94   \\
72     & 12     &  0.95  &  400     &  0.69   \\   
72     & 12     &  0.97  &  800     &  0.91   \\   
90     & 15     &  0.97  &  800     &  0.85   
\end{tabular}
\end{ruledtabular}
\end{table}

The viability of the underlying approach depends on whether it can actually produce
correctly structured helices. The first series of results measures the fraction of
chains that successfully fold into a helical state, and the manner in which the
success rate depends on $L$ and the cooling rate. A summary appears in
Table~\ref{tab:runs}; $L$ ranges from 18 to 90, which, since the helix period is
six, corresponds to 3--15 full helical turns.

Owing to the large number of runs it is not possible to provide a detailed history
of each, so a quantitative measure of folding success must be introduced. A
successfully folded helix is deemed to be one for which $S > 0.88$ at least once
during the last $1.2 \times 10^5$ steps of the run (measurements are made every
4000 steps); by this stage of the run the system has reached a comparatively low
temperature, so that further substantial conformational changes are unlikely.
Visual analysis confirms that, for the cases considered, this threshold for $S$
provides a quite reliable estimator; it tends to be sensitive to defects in the
helical structure, while allowing for the fact that a properly folded helix may
still has some residual curvature along its major axis.

It is clear from Table~\ref{tab:runs} that a high success rate for helix production
is achieved. Two trends are apparent in the results, neither of them unexpected:
for a given $f_T$, longer chains are less likely to fold properly than shorter
chains, and, for a given $L$, a larger $f_T$ (corresponding to slower cooling)
raises the success rate. Thus the longer the chain, the slower the desired cooling
rate; additional runs with faster cooling confirm this observation. The longest of
the chains folds to a helix with 15 turns, which, considering the potential for
defects, represents a significant victory of energy over entropy.

\begin{figure}
\includegraphics[scale=0.77]{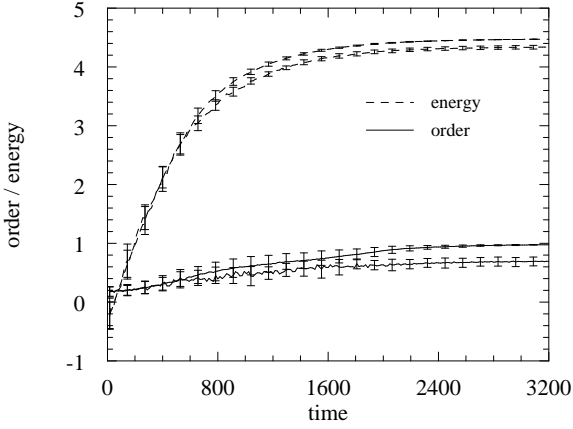}
\caption{\label{fig:evol54}Averaged order parameter and (negative) total energy per
DOF as functions of time (in dimensionless MD units) for chains with $L = 54$; the
contributions of chains that do and do not fold correctly appear in separate
curves, with the upper curve in each case corresponding to the successful folders.}
\end{figure}

\begin{figure}
\includegraphics[scale=0.77]{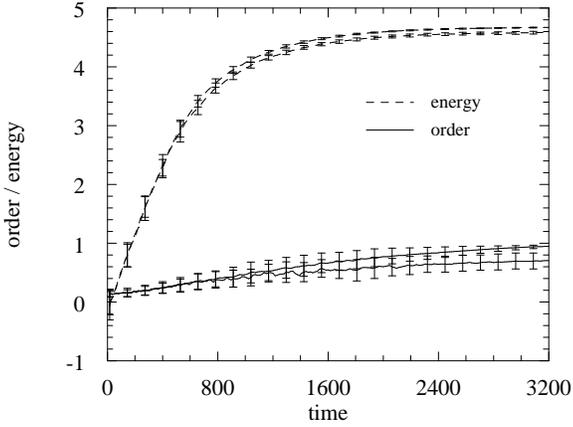}
\caption{\label{fig:evol90}Order parameter and energy for $L = 90$ (similar to
Fig.~\ref{fig:evol54}).}
\end{figure}

The rate at which chains approach the helically ordered state can be studied by
monitoring the mean values of $S$, as well as the negative of the total energy
(which is dominated by the torsional component when in the folded state); these
quantities provide measures of the long- and short-range order, respectively. The
results, normalized per DOF, for 54- and 90-site chains, averaged over all 400
runs, are shown in Figs.~\ref{fig:evol54} and \ref{fig:evol90}. The overall results
are divided into two groups, depending on whether the chain is classified as having
folded successfully or not, and error bars indicate the standard deviations of the
measurements. In each case it is the upper curve that represents the average for
the successfully folded chains, and it has the smaller error bars.

\begin{figure}
\includegraphics[scale=0.77]{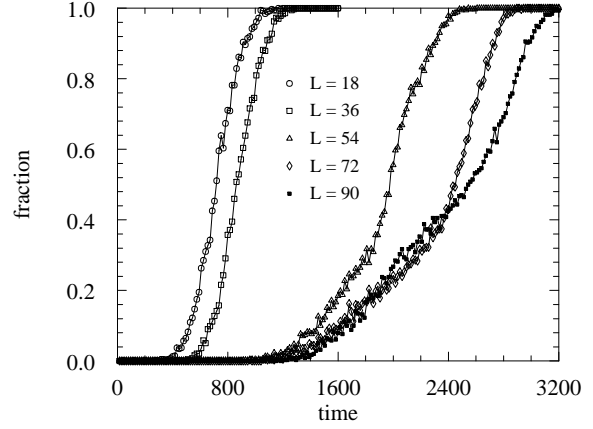}
\caption{\label{fig:cumltv}Cumulative distributions of chains in the folded state
as a function of time, for different lengths ($L$).}
\end{figure}

An alternative estimate of the rate at which folding proceeds is based on the time
dependence of the fraction of chains in a helical state, relative to all those that
eventually succeed in reaching this state. This provides information about when,
assuming a chain folds successfully, the appearance of helical order actually
occurs. Fig.~\ref{fig:cumltv} shows these cumulative distributions for the
different $L$, each at the slowest cooling rate considered. The cooling rate
clearly affects the results, as can be seen from the separation of the two groups
of curves that are based on different rates (see Table~\ref{tab:runs}); for a given
cooling rate, the folding speed tends to drop as the chains become longer (the
slight crossover of the $L = 72$ and 90 curves is probably not significant).

\begin{figure}
\includegraphics[scale=0.77]{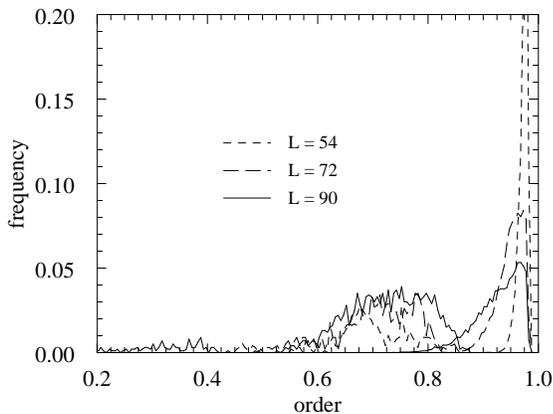}
\caption{\label{fig:sdistn}Order parameter distributions; separate curves show
results for chains that did (peaks on the right) and did not (multiplied by a
factor of 10) fold correctly.}
\end{figure}

A more detailed examination of final-state conformations is based on histograms of
the $S$-distribution, using measurements made over the last $1.2 \times 10^5$
steps. These results appear in Fig.~\ref{fig:sdistn} for several $L$ values (at the
slowest cooling rate). There are two separate curves for each $L$, one showing the
spread of $S$ for those chains that satisfied the folding criterion at least once
during this measurement period, and a broader, much lower curve for the chains that
did not. The former set of distributions become broader with increasing $L$; there
are several contributing causes for this, including slower folding rates (all the
runs were of equal length), chains not managing to fold successfully but having one
or more intermediate $S$ values which passed the test, and the increasing effect of
bending along the helical axis. The latter, broader distributions (scaled up by a
factor of 10 to make the details visible) are due both to the many defective
structures possible, each with its own spread of $S$ values, and also to the
defects themselves reducing the structural rigidity and so raising the
susceptibility to slow thermal vibration. Only for the longest chains is there any
overlap of the curves, and even then it is minimal.

The best way to follow the folding process is by viewing animated sequences of
images taken at various points during the run; some sequences can actually be
generated while running the simulation interactively, if the computations proceed
sufficiently rapidly. Here, due to the limitations of the printed page, a selection
of static images must suffice. One could attempt a verbal description of what
transpires but, as was the case in \cite{bvc98}, there are no obvious features
shared by the individual folding trajectories. Even if certain common
characteristics do exist, the strong random conformational fluctuations make their
observation difficult; a systematic, quantitative means for identifying pathways,
that extends ideas used for equilibrium states \cite{cri98}, might prove helpful in
this task.

\begin{figure}
\includegraphics[scale=0.38]{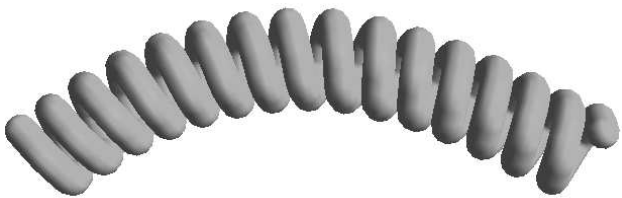}
\caption{\label{fig:lk3d_02}Correctly folded helix ($L = 90$) with residual
curvature; this is the most extreme case of bending observed.}
\end{figure}

Figure~\ref{fig:lk3d_01}, which appeared early in the paper, shows an image of a
typical, well-formed, almost straight helix obtained in one of the runs, while
Fig.~\ref{fig:lk3d_01a} shows a random chain configuration observed near the start
of a run, both for $L = 90$ chains. For clarity, these and subsequent pictures
represent the chains by their tubular envelopes, rather than by showing individual,
partially overlapped spheres representing the monomers. The tube thickness
corresponds to unit diameter, slightly less than the soft-sphere interaction cutoff
to ensure that a small amount of space remains visible between adjacent turns of
the helix. The maximum amount of residual curvature that was observed in the
backbones of properly folded helices is apparent from Fig.~\ref{fig:lk3d_02}; this
example actually meets the criterion for folding success (as indeed it should).

\begin{figure}
\includegraphics[scale=0.38]{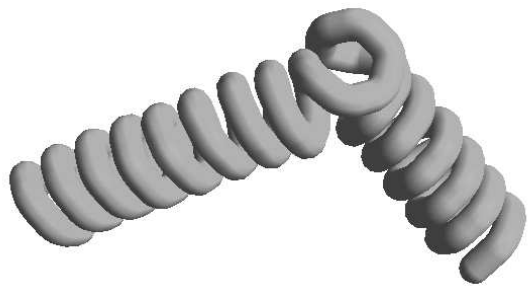}
\caption{\label{fig:lk3d_03}Incorrectly folded state with a single defect.}
\end{figure}

While the majority of runs (85\% for chains with $L = 90$, and an even higher
proportion for smaller $L$, see Table~\ref{tab:runs}) result in a correctly folded
helix, examination of the kinds of defects that appear in the final states of those
runs that fail to fold properly is an informative exercise. The first such picture,
Fig.~\ref{fig:lk3d_03}, is of an $L = 90$ chain with two helical regions separated
by a single defect. The defect is essentially a single loop of the helix with a
reverse fold that became frozen in place during the cooling process. This is the
most frequent type of defect, and its location can be anywhere in the chain, even
right at the end.

\begin{figure}
\includegraphics[scale=0.38]{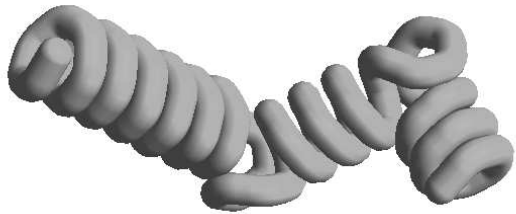}
\caption{\label{fig:lk3d_04}Incorrectly folded state with two separate defects.}
\end{figure}

\begin{figure}
\includegraphics[scale=0.4]{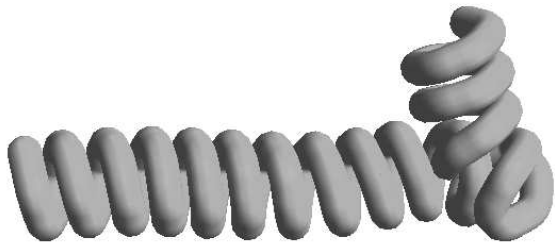}
\caption{\label{fig:lk3d_05}Folded state with a localized intertwined defect.}
\end{figure}

\begin{figure}
\includegraphics[scale=0.4]{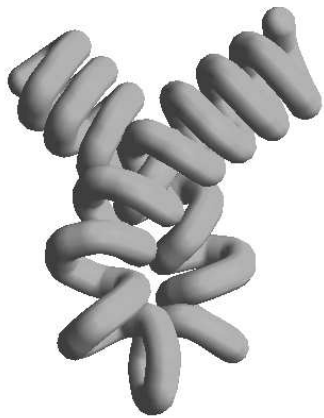}
\caption{\label{fig:lk3d_06}Folded state with a complex defect involving a large
loop.}
\end{figure}

Less frequent are chains with two spatially separated defects, as shown in
Fig.~\ref{fig:lk3d_04}. More extreme, but very rare examples of other kinds of
defects appear in Figs.~\ref{fig:lk3d_05} and \ref{fig:lk3d_06}. These show what
can happen when the chain starts to become entangled with itself; in the first case
the problem is localized, but in the second (the only example of its kind observed)
there is a relatively large loop trapped by the entanglement. The fact that these
defects are relatively infrequent, and that even this low level of failure can be
reduced by lowering the cooling rate still further, attest to the robustness and
reliability of the folding process.

\subsection{Folding to a pair of helices}

The helix formation described above is obtained in a study of homopolymers where,
due to the uniformity of the chain, the only kind of repeating structure that can
be produced is the helix (the planar zigzag conformation is a degenerate case). In
a protein context this kind of structure is classed as secondary because helices
often serve as structural components in more complex assemblies. Globular proteins
are characterized by at least one additional level in the structural hierarchy,
namely tertiary structure. A model capable of demonstrating tertiary structure
requires at least some differentiation among the chain sites that breaks the
translational invariance. Building on the helix-forming model investigated here,
the next stage of complexity is a packed assembly of helices, a structure that
incorporates both the secondary and tertiary levels of the hierarchy. The simplest
way to design such a structure is to include non-local, attractive forces between
selected pairs of chain sites; here, the pairs involved are located at chain
positions that will be brought into proximity following the collapse into a state
with two adjacent helices aligned in antiparallel directions. Such highly specific
interactions are reminiscent of an approach used for lattice protein models
\cite{ued78}; the overall simplicity should be contrasted with the highly detailed
model, complete with solvent, used in an MD study of the {\em un}folding of a
three-helix bundle \cite{boc95}.

Choosing the interactions to produce a twin-helix structure is accomplished as
follows: For a homogeneous chain with $L = n_b+1$ sites, in which the periodicity
of the helix is $p$, the ground state consists of $n_t = L/p$ turns. Now assume
that $n_t$ is an odd number and choose the interactions appropriate for a pair of
adjacent helices, each with $(n_t-1)/2$ turns, joined by a ``bridging'' chain
segment of length $p$. All that remains is to identify the pairs of sites in the
two helical regions that must attract; these are just neighboring sites in adjacent
turns of one of the helical segments, matched with the corresponding sites, in
reverse order, of the other. The strength of the attractive potential responsible
for the tertiary structure, which is based on Eq.~(\ref{eq:sspot}), is $0.2
u^{(0)}$; it is weaker than the torsion, but the question of whether tertiary
structure formation is the beneficiary or the cause of secondary structure
formation \cite{dil95} is not addressed here. In these exploratory computations,
the torsional interactions along the bonds in the bridging segment are set to zero
for simplicity; such interactions could actually be used to assist the folding and
will be the subject of future study.

\begin{figure}
\includegraphics[scale=0.77]{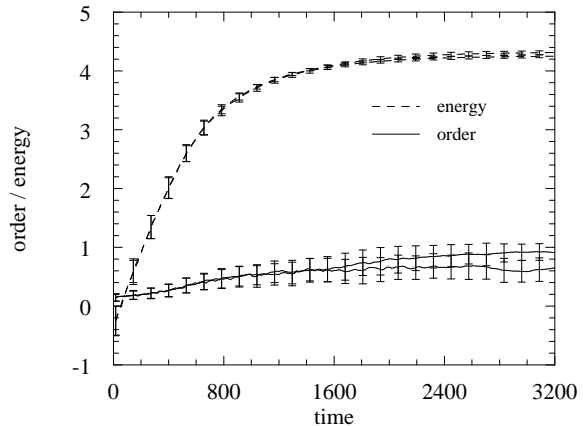}
\caption{\label{fig:evol78}Averaged order parameter and energy as functions of time
for chains that form antiparallel helix pairs ($L = 78$).}
\end{figure}

The definition of the long-range order parameter $S$, Eq.~(\ref{eq:sdef}), must be
modified to reflect the structure of the anticipated collapsed state. The partial
contributions to $S$ of the two helical segments are now combined with opposite
signs, and contributions from the bridging region ignored; this provides a
reasonably sensitive, but unambiguous, measure of folding success.
Fig.~\ref{fig:evol78} shows how both the modified $S$ and the energy vary with
time, for $L = 78$ chains (corresponding to a folded state consisting of a pair of
6-turn helices), using runs whose details are otherwise similar to those for $L =
90$. Based on visual analysis of the behavior, a different definition of what
constitutes successful folding is needed here, namely that the value of the
modified $S$ must now exceed 0.95 for folding to be considered successful; using
this criterion the success fraction was found to be 0.66.

\begin{figure}
\includegraphics[scale=0.4]{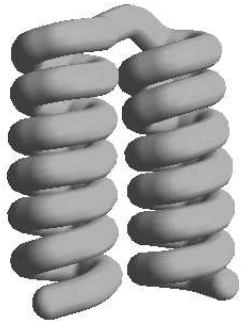}
\caption{\label{fig:lk3d_07}Well-formed pair of helices with antiparallel
alignment.}
\end{figure}

\begin{figure}
\includegraphics[scale=0.4]{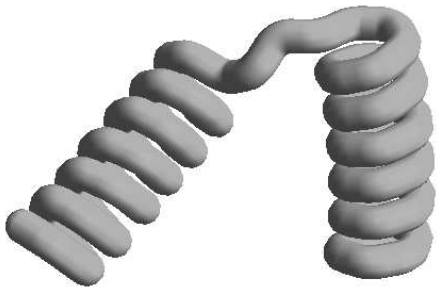}
\caption{\label{fig:lk3d_08}Pair of helices that have failed to align.}
\end{figure}

Figure~\ref{fig:lk3d_07} shows an example of a successfully folded helix pair, an
ordered conformation in which both secondary and tertiary structural elements are
manifest; two-thirds of the runs ended in this state. The failure to fold properly
was generally not due to defects in the individual helical segments, but because
the two secondary components did not succeed in aligning correctly; an example of
such an outcome is shown in Fig.~\ref{fig:lk3d_08}. The attractive forces become
much more effective once the helical segments have formed; this is due to the
linear arrangement of the attraction sites enabling them to function cooperatively,
an effect that may be reflected in real proteins with prominent
secondary-structural features. As a result of the nature of the interactions and
the relative interaction strengths, the helical segments form first, essentially
unimpeded, and only then do they attempt to align. The failure to align here is a
symptom of the absence of any driving force for bringing the helices together;
there is no torsional preference in the bridging segment and the range of the
inter-helix attraction is too short to be felt if the helical segments are well
separated. Changes to either or both these aspects of the potential should alter
the behavior, but care is required to avoid hindering the helix formation process
in any way.

\section{Conclusions}

The present paper has focused on both methodology and results. A formalism
developed for the dynamics of robotic manipulators and other coupled mechanical
systems -- that provides a convenient and direct representation of the dynamics of
bodies connected by rigid links with restricted degrees of freedom -- has been
utilized in a polymer context, with the clear implication that existing methods
based on geometric constraints may be redundant in many instances. Since the
treatment also involves dealing with rigid-body dynamics, a computationally more
effective method than the often-used quaternion approach is also employed.

The results of an extensive series of MD simulations demonstrate that homopolymer
chains with suitable torsional interactions consistently collapse into well-formed
helices; the probability of localized defects being frozen into the structure
depends on the cooling rate, and it can be reduced to a very low level by cooling
sufficiently slowly. In order to demonstrate that the present simplified approach
is relevant for protein folding, heterogeneous chains, with interactions favoring
the development of antiparallel pairs of helices, were shown to produce coexisting
secondary and tertiary structural features.

The order parameters introduced to quantify the degree of folding were tailored to
capture the structural order present in the final state of the polymer. To study
the details of folding pathways, other order parameters (or reaction coordinates)
that capture features present in the intermediate states, but not necessarily in
the final state, could be defined. In the twin-helix case, for example, a simple
sum of the absolute values of $S$, evaluated separately for each helix-forming
segment of the chain, might prove useful, since this quantity reaches its maximum
upon completion of secondary structure formation, and is not seriously affected by
subsequent rearrangement at the tertiary level.

The apparent success of the MD approach to chain folding used here is important for
another reason. The widely-cited Levinthal ``paradox'' \cite{pan00} implies that
since the number of states accessible to a protein grows exponentially with residue
count, the time required for even a small protein to seek out its native state is,
for practical purposes, infinite. Since nature does not suffer from this problem,
the implication is that substantial portions of the folding process occur along
certain well-characterized pathways; thus the molecules do not really wander almost
aimlessly through conformation space, and hence there is no paradox. In order to
begin to simulate such processes it is necessary to resort to a computationally
efficient model, with realistic dynamics and a unique but readily determined
low-energy ``native'' state; this is precisely what has been accomplished in the
present work.

The type of model introduced here provides a starting point for exploration in
several directions. While the interactions were weighted to construct the secondary
helix structure prior to forming features at the tertiary level, a change in the
relative strength of the interactions would allow aspects of both levels of
organization to appear concurrently. Chains could be designed to fold into other
idealized compact structures, such as the packed cube used in some lattice studies
\cite{dil95,sha97}, or sheetlike conformations that also represent important
secondary structure components; furthermore, packed states with different degrees
of accessibility could provide useful information on how this feature influences
folding success. The interactions can be modified and new types of interactions
added; polymer topology can be changed by the addition of side chains corresponding
to residues with extended structure. Common to all these enhancements is that the
model must always be designed with a known lowest-energy state, and in this respect
the approach differs from many other types of protein simulation. While models of
this kind are perhaps limited in the kinds of questions they can address, there are
more than enough issues requiring attention where they can prove helpful.

In a sense, the role played by such highly simplified models is analogous to the
Ising model of ferromagnetism \cite{hua63}; while it is not usually claimed that an
Ising spin system accurately represents a real magnetic material (or, for that
matter, any other kind of real physical system, when used for other kinds of
problems such as lattice gases) it does however capture a great deal of the essence
of the problem, to an extent that the study of Ising and related models has
resulted in important advances, both for spin systems in particular, and for
statistical mechanics and critical phenomena in general. Proteins can also be
modeled with a high level of detail and specificity, but the tradeoff is that only
short trajectory segments can be followed with an investment of a reasonable amount
of computing effort; hopefully, extensive studies of simplified polymer models of
the kind examined here, in which the design is tailored to reproduce certain
generic aspects of macromolecular behavior, will achieve greater popularity as
their usefulness becomes established.

\begin{acknowledgments}
This work was partially supported by the Israel Science Foundation.
\end{acknowledgments}

\bibliography{chainhelix}

\end{document}